\documentclass{elsart}
\usepackage{mathrsfs}
\usepackage{amssymb}
\usepackage{bm}
\usepackage{epsfig,eepic,epic}
\usepackage{graphicx}
\usepackage[dvips]{color}
\usepackage[hypertex,linkcolor=red]{hyperref}
\newcommand{\kk}{\rangle\!\rangle}\newcommand{\bb}{\<\!\<}
\newcommand{\Tr}{{\rm Tr}}\def\<{\langle}\def\>{\rangle}
\def\dim{\operatorname{dim}}\def\d{\operatorname{d}}
\def\sH{{\mathcal H}}\def\Reals{\mathbb R}
\begin{document}
\begin{frontmatter}
 \title{On the realization of Bell observables}
\author{Giacomo Mauro D'Ariano and Paolo Perinotti}
\address{QUIT Group, Dipartimento di Fisica A. Volta, Universit\`a di Pavia\\
and Istituto nazionale di Fisica della Materia\\ via
  Bassi 6, I-27100 Pavia, ITALY\\ {\tt http://www.qubit.it}
}\date{\today}
\begin{abstract}
  We show how Bell observables on a bipartite quantum system can be
  obtained by local observables via a controlled-unitary
  transformation. For continuous variables this result holds for the Bell observable corresponding
  to the non-conventional heterodyne measurement on two radiation modes, 
which is connected 
through a 50-50  beam-splitter to   two local observables given by single-mode homodyne
measurements. A simple scheme for a controlled-unitary transformation of continuous variables is
also presented, which needs only two squeezers, a parametric down-converter and two beam
  splitters.
\end{abstract}
\end{frontmatter}
\section{Introduction}
The non-classical correlations of entangled quantum systems are the
basis for the engineering of the next-generation devices for quantum
information processing\cite{Nielsen2000}.  In particular, the
so-called {\em Bell measurements} play a pivotal role in most quantum
processing techniques, as they are essential in any teleportation
scheme\cite{bbjcpw} or dense coding protocol\cite{dens}, and recently
have been proved an invaluable resource for achieving "informationally
complete" measurements\cite{infocompl}. Nonetheless the Bell
measurements still represent a serious experimental challenge, and a
general scheme for designing them would be particularly welcomed.
\par The name {\em Bell measurement} is generally designating 
a non separable joint measurement on a bipartite quantum system,
typically a POVM made of rank-one operators proportional to projectors
on maximally entangled vectors \cite{telep}. Here we will focus
attention on the special case of the {\em Bell observable},
corresponding to an orthonormal basis of maximally entangled vectors.
We will show how a Bell observable can be achieved by local
measurements via a (nonlocal) interaction of the {\em
  controlled-unitary} form---a generalization to dimension $d>2$ of the controlled-NOT for
qubits---corresponding to a {\em coherence-preserving choice} among unitary 
transformations controlled by the state preparation of an ancilla. This
result is interesting because it emphasizes the pivotal role of the
controlled-unitary transformation in quantum information processing.
For continuous variables the same result holds for the Bell observable realized by heterodyning  
two radiation modes, which is connected through a 50-50 beam-splitter to two local observable
describing single-mode homodyne measurements. In this way we
will also see how a simple scheme for a controlled-unitary
transformation of continuous variables can be achieved, by just using
only two squeezers, a parametric down-converter, and two beam
splitters.\par

After introducing some useful notation in Section \ref{notat}, in
Section \ref{findim} we prove the connection between Bell and local
observables for finite dimension. In Section \ref{beamsplitt} we give
the continuous-variable case, connecting the heterodyne Bell
observable to single-mode homodyne measurements via a 50-50
beam-splitter, and presenting the optical scheme for the
controlled-unitary transformation. Section \ref{conclusion} closes the
paper with a summary and open problems.
\section{Some notation}\label{notat}
We will make extensive use of the following correspondence between
Hilbert-Schmidt operators on a the Hilbert space $\sH$ and vectors in
$\sH \otimes\sH $ \cite{bellobs}
\begin{equation}
A=\sum_{m,n}A_{mn}|m\>\< n|\,,\quad|A\kk=\sum_{m,n}A_{mn}|m\>|n\>\,,
\label{eq:hs}
\end{equation}
where the double-ket symbol $|A\kk$ will be used to remind the
correspondence of the vector $|A\kk\in\sH \otimes\sH $ with the
operator $A$ on $\sH $. The scalar product in $\sH \otimes\sH $
corresponds to the Hilbert-Schmidt scalar product between operators
\begin{equation}
\bb A|B\kk=\Tr[A^\dag B]\,,
\label{eq:scalp}
\end{equation}
and analogously the norm of vectors corresponds to the Frobenius norm.
The following identities will be handy for calculations
\begin{eqnarray}
A\otimes B|C\kk&=&|ACB^T\kk\,,\label{eq:prop1}\\
\Tr_1[|A\kk\bb B|]&=&A^TB^*\,,\label{eq:prop2}\\
\Tr_2[|A\kk\bb B|]&=&AB^\dag\,,\label{eq:prop3}
\end{eqnarray}
where $A^T$ and $A^*$ denote the transposed operator and the complex
conjugated operator of $A$, respectively, with respect to the basis used in
Eq. (\ref{eq:hs}).
\par From Eqs.  (\ref{eq:prop2}) and (\ref{eq:prop3}) it is clear that
for finite dimension $d=\dim(\sH)$ the maximally entangled states are
those corresponding to unitary operators scaled by $\frac1{\sqrt{d}}$,
since these are the only pure states having maximally chaotic local
states.  For infinite dimensions we will consider Dirac-normalizable
maximally entangled vectors.

\section{Systems with finite dimensional Hilbert space}\label{findim}
Let's consider the set of $d^2$ maximally entangled states
$d^{-\frac{1}{2}}|U(m,n)\kk$, corresponding to the {\em
  shift-and-multiply} unitary operators
\begin{equation}
U(m,n)=Z^m W^n\,,\label{Unm}
\end{equation}
where 
\begin{equation}
Z=\sum_{j=0}^{d-1}e^{\frac{2\pi i}{d}j}|j\>\< j|,\qquad
W=\sum_{j=0}^{d-1}|j\oplus 1\>\< j|,\label{ZW}
\end{equation}
where $\oplus$ denotes sum modulo $d$.  The operators $U(m,n)$ provide
a projective irreducible representation of the abelian group ${\mathbb
  Z}_d\times{\mathbb Z}_d$. It is easy to check \cite{bellobs} that
the operators $U(m,n)$ are an orthonormal basis for $\sH \otimes\sH $,
whence the vectors $d^{-\frac{1}{2}}|U(m,n)\kk$ are a Bell basis, i.
e. a maximally entangled orthonormal basis. This is precisely the
basis used for quantum teleportation in Refs. \cite{bbjcpw,telep}. For
qubits ($d=2$) it is the basis of Pauli matrices
$\{\frac1{\sqrt2}|I\kk\,,\frac1{\sqrt2}|\sigma_x\kk\,,
\frac1{\sqrt2}|\sigma_y\kk\,,\frac1{\sqrt2}|\sigma_z\kk\}$.
\par Our aim is now to find a unitary operator $V$ on $\sH\otimes\sH$
evolving a local basis into the Bell basis, more precisely such that
\begin{equation}
V|e_m,n\kk=\frac1{\sqrt d}|U(m,n)\kk,
\end{equation}
where as a local basis we choose $|e_m\>\otimes|n\>$, the vector
$|e_j\>$ denoting the Fourier transformed vector of $|j\>$, with
\begin{equation}
|e_j\>\equiv\frac1{\sqrt d}\sum_{n=0}^{d-1}e^{\frac{2\pi i}{d}nj}|n\>=F|j\>\,,\qquad
F\equiv\frac1{\sqrt d}\sum_{n=0}^{d-1}e^{\frac{2\pi i}{d}nj}|n\>\< j|\,.
\end{equation}
In the following we will often use the short notation
$|\phi,\psi\kk\doteq|\phi\>\otimes|\psi\>$ for tensor products of
vectors.

A formal expression for $V$ is readily given since we know its action
on a complete orthonormal set
\begin{equation}
V=\frac{1}{\sqrt{d}}\sum_{m,n=0}^{d-1}|U(m,n)\kk\bb e_m,n|\\=
\frac1{\sqrt
  d}\sum_{m,n=0}^{d-1}\sum_{i,j=0}^{d-1}U(m,n)_{ij}|i,j\kk\bb e_m,n|, 
\label{eq:formesv}
\end{equation}
where, using Eqs. (\ref{Unm}) and (\ref{ZW}) we recover the matrix
elements $U(m,n)_{ij}$ as follows
\begin{equation}
U(m,n)_{ij}\doteq \bb i,j|U(m,n)\kk
=e^{\frac{2\pi i}{d}im}\delta_{i\oplus n,j}\,.
\end{equation}
This corresponds to the following expression for $V$
\begin{equation}
V=\frac1{\sqrt d}\sum_{m,n,i=0}^{d-1}e^{\frac{2\pi i}{d}im}|i\>\< e_m|\otimes|n\oplus i\>\< n|=
\frac1{\sqrt d}\sum_{n,i=0}^{d-1}|i\>\< i|\otimes|n\oplus i\>\< n|\,.
\end{equation}
Thus we have
\begin{equation}
V=\sum_{i=0}^{d-1}|i\>\< i|\otimes W^{i}\,,
\label{eq:contru}
\end{equation}
namely the unitary $V$ is a {\em controlled-unitary} transformation,
corresponding to choosing (coherently) among the unitary
transformations $W^{i}$ via state preparation of the first system in
$\sH\otimes\sH$. The transformation in Eq. (\ref{eq:contru})
generalizes to dimension $d>2$ the well known controlled-NOT gate for
qubits.  The present result is also in agreement with a similar one
implicit in Ref. \cite{generbell}.\par 

In the next section we will see how the present controlled-unitary
evolution can be generalized to infinite dimensions, and how the
local-Bell connection is achieved by a very common device: the beam
splitter.

\section{Continuous variables}\label{beamsplitt}
The derivation of Section \ref{findim} cannot be generalized
straightforwardly, since the group ${\mathbb Z}_d\times{\mathbb Z}_d$
has no extension to $d=\infty$. However, we will see how the same
construction can be carried out for "continuous variables"---i. e. for
continuous spectrum and bosonic modes---corresponding to an actual
quantum optical implementation.\par Let's consider two bosonic modes,
with creation and annihilation operators denoted by $a^\dag$,
$b^\dag$, and $a$, $b$, respectively. In the Hilbert space of each
mode consider the complete sets of Dirac-normalized eigenvectors
$\{|x\>_0\}$ and $\{|x\>_{\frac\pi2}\}$ of the quadratures
$X_0=\frac{1}{2}(a^\dag+a)$ and $X_{\frac\pi2}=\frac{i}{2}
(b^\dag-b)$.  Consider the following unitary operator
\begin{equation}
C=\frac1{\sqrt{\pi}}\int_\Reals\d x\int_\Reals \d
y\,|\Delta(x,y)\kk _0\<x|\otimes_{\frac\pi2}\<y|,
\label{eq:contrunop}
\end{equation}
where $\Delta(x,y)\equiv e^{-2 i x X_\frac\pi2}
e^{2iyX_0}=e^{-ixy}D(x+iy)$, $D(\alpha)\doteq e^{\alpha
  a^\dag-\alpha^* a}$ denoting the displacement operator. Notice that
in the present infinite-dimensional setting the notation $|A\kk$ in
Eq. (\ref{eq:hs}) corresponds to $|A\kk\doteq A\otimes |I\kk$ with
$|I\kk$ denoting the generalized vector
$|I\kk=\sum_{n=0}^\infty|n\>\otimes|n\>\equiv \int_\Reals\d x
|x\>_0\otimes |x\>_0$, where in the representation of eigenstates of
$a^\dag a$ the transposition of mode operators is given by
$a^T=a^\dag$.
\par The operator $C$ can be considered as the infinite
dimensional analog of the operator $V$ in Eq.  (\ref{eq:formesv}),
with the orthonormal basis $\{|m\>\}$ replaced by the continuous Dirac
set $\{|x\>_0\}$. The basis $\frac1{\sqrt\pi}|\Delta(x,y)\kk$ is
indeed maximally entangled, with orthogonality relations
$\bb\Delta(x,y)|\Delta(x',y')\kk= \pi\delta(x-x')\delta(y-y')$. The
following calculation proves that $C$ is a controlled-unitary
transformation
\begin{eqnarray}
C=
&&\frac1{\sqrt{\pi}}\int_\Reals\d x\int_\Reals \d
y\int_\Reals \d t\,e^{-2 i x X_\frac\pi2}\otimes e^{2iyX_0}|t\>_{00}\<x|\otimes|t\>_{0\frac\pi2}\<y|=\nonumber\\
&&\frac1{\sqrt{\pi}}\int_\Reals\d x\int_\Reals \d
y\int_\Reals \d t\,(I \otimes e^{2iyt})
|t+x\>_{00}\<x|\otimes|t\>_{0\frac\pi2}\<y|=\nonumber\\
&&\frac1{\sqrt{\pi}}\int_\Reals \d y\int_\Reals \d t\,e^{-2 i t X_\frac\pi2}\otimes |t\>_{0\frac\pi2}\<y|e^{2iyt}=\nonumber\\
&&\left(\int_\Reals \d t\,e^{-2 i t X_\frac\pi2}\otimes
  |t\>_0{}_0\<t|\right)\left( I\otimes\int_\Reals\d
  y \,|y\>_{\frac\pi2\frac\pi2}\<y|\right) 
=\nonumber\\
&&\int_\Reals \d t\,e^{-2 i t X_\frac\pi2}\otimes
  |t\>_0{}_0\<t|\,.
\end{eqnarray}
Therefore, also for continuous variables we have a unitary operator
that maximally entangles a factorized basis, and which corresponds to
a controlled-unitary transformation.

\subsection{Achieving the controlled-unitary $C$}

In this section we will give an optical scheme for achieving $C$, involving
the use of parametric down-conversion and beam splitters.
\par
\begin{figure}[h]
\input{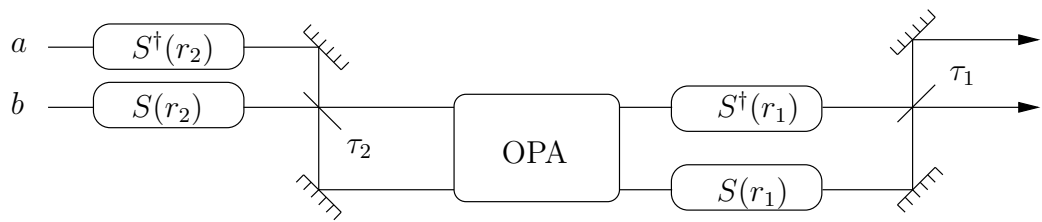}
\label{fig:scheme}
\caption{Optical scheme to achieve the controlled unitary transformation
  (\ref{eq:contrfin}) (Schr\H{o}dinger picture). $S(r)$ represents the squeezing transformation
  $R(r)\doteq e^{\frac12\log r ({a^\dag}^2-a^2)}$. The values of the
  squeezing parameters are $r_1=2^{-\frac{1}{2}}$ and
  $r_2=(\frac{3}{4})^{-\frac{1}{4}}$. The beam splitters have
  transmissivities $\tau_1=\{4(2-\sqrt3)\}^{-1}$ and $1/2$. The
  optical parametric amplifier (OPA) has field-amplitude gain
  $g=\{2(3-2\sqrt3)\}^{-1}$.}
\end{figure}

 For a suitable choice of phases of the input
and output modes, a 50-50 beam-splitter can be represented by the
unitary operator
\begin{equation}
V=e^{\frac\pi4(a^\dag b-ab^\dag)}\,.
\label{eq:beamsplitt}
\end{equation}
The above unitary operator brings a local basis of two homodyne
detectors into the Bell basis of a heterodyne detector \cite{bilkent}.
In fact, consider the joint homodyne detection of two "orthogonal"
quadratures, described by the generalized eigenvectors
$|x\>_0\otimes|y\>_{\frac\pi2}$ of the quadratures $X_0\otimes
X_{\frac\pi2}$. It is immediate to show that $|D(z)\kk$ is the
generalized eigenvector of the heterodyne photocurrent $Z=a-b^\dag$
corresponding to the complex eigenvalue $z$. We will now prove the
following relation
\begin{equation}
V\left|\frac x{\sqrt 2}\right\>_0\otimes\left|\frac y{\sqrt
    2}\right\>_{\frac\pi2}=\left(\frac2\pi\right)^\frac12|D(x+ i y)\kk\,.\label{eq:entbs}
\end{equation}
First remind that the generalized eigenvectors $|x\>_\phi$ of the
generic quadrature $X_\phi=\frac12(e^{ i \phi}a^\dag+e^{- i \phi}a)$
can be written as
\begin{equation}
|x\>_\phi=e^{- i \phi a^\dag a}D(x)|0\>_0=\left(\frac2\pi\right)^\frac14e^{- i \phi a^\dag
  a}D(x)e^{-\frac{a^\dag{}^2}2}|0\>\,,\label{vacuumx}
\end{equation}
where $|0\>$ denotes the vacuum for mode $a$. Upon rewriting
$|x/\sqrt2\>_0\otimes|y/\sqrt2\>_{\frac\pi2}$ via identity
(\ref{vacuumx}) in terms of operators acting on the vacuum, we apply
the beam splitter operator $V$ on the left, and after some algebra we
obtain
\begin{equation}
V\left|\frac x{\sqrt 2}\right\>_0\otimes\left|\frac y{\sqrt
    2}\right\>_{\frac\pi2}=\left(\frac2\pi\right)^\frac12e^{-\frac12(x^2+y^2)+a^\dag(x+ i y)-b^\dag(x- i y)}e^{a^\dag b^\dag}|0\>\otimes|0\>\,.
\end{equation}
Notice that $e^{a^\dag b^\dag}|0\>\otimes|0\>=|I\kk$. Eq.
(\ref{eq:entbs}) then follows immediately by using Eq.
(\ref{eq:prop1}). We are now ready to derive the optical scheme for
the controlled-unitary $C$. According to Eq.  (\ref{eq:entbs}), we can
rewrite $V$ in the form
\begin{equation}
V(I\otimes e^{-i \frac\pi2b^\dag b})\left|\frac x{\sqrt 2}\right\>_0\otimes\left|\frac y{\sqrt 2}\right\>_0
=\left(\frac2\pi\right)^\frac12e^{i xy}|\Delta(x,y)\kk\,.
\label{Delta}
\end{equation}
Introducing the local squeezing transformation
\begin{equation}
S(r)|x\>=\sqrt{r}\left|r x\right\>\,,\label{R}
\end{equation}
we can rewrite Eq. (\ref{Delta}) as follows
\begin{eqnarray}
V(I\otimes  e^{-i \frac\pi2b^\dag b})\left[S\left(2^{-\frac{1}{2}}\right)\otimes S\left(2^{-\frac{1}{2}}\right)\right]
e^{- i X_0\otimes X_0}(I\otimes  e^{i\frac\pi2b^\dag b})=\nonumber\\
\frac1{\sqrt{\pi}}\int_\Reals\d x\int_\Reals \d
y\,|\Delta(x,y)\kk_0\< x|_{\frac\pi2}\<y|\,.\label{eq:convarcontru} 
\end{eqnarray}
A realization of the transformation (\ref{R}) is given by the unitary
single-mode squeezing operator
\begin{equation}
S(r)=e^{\frac12\log r({a^\dag}^2-a^2)}.\label{eq:contrun}
\end{equation}
In l.h.s. of Eq. (\ref{eq:convarcontru}) the only unitary operator
that has no direct physical interpretation is the exponential $e^{- i
  X_0\otimes X_0}$. We will now write it as a product of physical
unitary transformations. First, let us write the exponential in terms
of bosonic operators
\begin{equation}
e^{- i X_0\otimes X_0}=e^{-\frac i 4(a^\dag+a)(b^\dag+b)}=e^{-\frac i 4(a^\dag b^\dag+ab+ab^\dag+a^\dag b)}\,.
\end{equation}
The operators $K_x=\frac12(a^\dag b^\dag+ab)$ and $K_y=\frac i
2(ab^\dag+a^\dag b)$ along with their commutator
\begin{equation}
K_z=i [K_x,K_y]=\frac14({a^\dag}^2-a^2+{b^\dag}^2-b^2)
\label{eq:comm}
\end{equation}
are the generators of the Lie algebra $su(1,1)$. In terms of the
generators of the Lie algebra, the exponential $e^{- i X_0\otimes
  X_0}$ is simply given by
\begin{equation}
e^{- i X_0\otimes X_0}=e^{-\frac i 2(K_x- i K_y)}=e^{-\frac i 2K_-}\,.
\label{eq:e-i2k-}
\end{equation}
Using the Pauli matrix realization of the Lie algebra $K_{x/y}=\frac i
2\sigma_{x/y}$ and $K_z=\frac12\sigma_z$, one can easily derive the
identity
\begin{equation}
e^{-\frac i 2K_-}=e^{ i \alpha K_x}e^{\beta K_y}e^{\gamma K_z}\,,
\end{equation}
where
\begin{equation}
\alpha=-2\tanh^{-1}(2-\sqrt3)\,,\quad
\beta=-2\tan^{-1}(2-\sqrt3)\,,\quad
\gamma=\log\frac{\sqrt3}2\,,
\end{equation}
namely one has
\begin{equation}
e^{-iX_0\otimes X_0}=e^{\frac i 2\alpha(a^\dag b^\dag+ab)}e^{\frac i 2\beta(a^\dag b+ab^\dag)}e^{\frac14\gamma({a^\dag}^2-a^2+{b^\dag}^2-b^2)}\,.
\end{equation}
Summarizing the above results, the controlled-unitary transformation
$C$ is realized as follows
\begin{equation}
C=V(S(r_1)\otimes S(r_1)^\dag)e^{-\frac12\alpha(a^\dag b^\dag-ab)}e^{\frac12\beta(a^\dag b-ab^\dag)}(S(r_2)^\dag\otimes S(r_2))\,,
\label{eq:contrfin}
\end{equation}
with $r_1=2^{-\frac{1}{2}}$ and $r_2=(\frac{3}{4})^{-\frac{1}{4}}$.
The overall scheme for the controlled unitary transformation is given
in Fig. \ref{fig:scheme}.
\section{Conclusions}\label{conclusion}
We have shown how a Bell measurement can be obtained from local
measurements via an interaction of the controlled-unitary form. The
considered Bell measurement is the one used in the original teleportation
proposal of Ref. \cite{bbjcpw}. We conjecture that our result holds
more generally for every Bell measurement: however, a general proof would
require a complete classification of all Bell measurements, which by
itself is still an open problem. We have seen that for continuous
variables a simple 50-50 beam splitter achieves the factorization of
the heterodyne Bell measurement, whereas a controlled-unitary
interaction can be achieved by means of two squeezers, a parametric
down-converter and two beam splitters.  This result can be certainly
of interest for applications to continuous-variable processing of
quantum information.
\section*{Acknowledgments}
This work has been sponsored by INFM through the project
PRA-2002-CLON, and by EEC and MIUR through the cosponsored ATESIT
project IST-2000-29681 and Cofinanziamento 2003.

\end{document}